\documentclass[preprint]{JHEP3} 

\usepackage{epsfig,multicol}

\newcommand\fverb{\setbox\pippobox=\hbox\bgroup\verb}
\newcommand\fverbdo{\egroup\medskip\noindent%
			\fbox{\unhbox\pippobox}\ }
\newcommand\fverbit{\egroup\item[\fbox{\unhbox\pippobox}]}
\newbox\pippobox

\newcommand{\be}{\begin{equation}} 
\newcommand{\ee}{\end{equation}}

\newcommand{\ba}{\begin{eqnarray}}
\newcommand{\ea}{\end{eqnarray}}

\title{D0-brane tension in string field theory}

\author{Matteo Beccaria\thanks{Partially supported by INFN, IS-LE21}\\
	Dipartimento di Fisica, Universita' di Lecce,\\
        Via Arnesano, 73100 Lecce\\
	INFN, Sezione di Lecce\\
	E-mail: \email{matteo.beccaria@le.infn.it}}

\abstract{
We compute the D0-brane tension in string field theory by representing it as a tachyon lump 
of the D1-brane compactified on a circle of radius $R$. 
To this aim, we calculate the lump solution in level truncation up to level $L=8$.
The normalized D0-brane tension is independent on $R$. The compactification radius is 
therefore chosen in order to cancel the subleading correction $1/L^2$.
We show that an optimal radius $R^*$ indeed exists and that at $R^*$ the theoretical
prediction for the tension is reproduced at the level of $10^{-5}$. 
As a byproduct of our calculation we also discuss the determination of the 
marginal tachyon field at $R\to 1$.
}


\begin{document} 


\section{Introduction}
\label{Sec:Intro}

The critical bosonic open string admits Dp-branes for all $p\le 26$, and each brane has a tachyonic mode
with squared mass $M^2 = -1$ in units where the string tension is $\alpha' = 1$. 
Since 1999, Ashoke Sen proposed in several steps important new insights into the  non perturbative dynamics of string theory~\cite{Sen:1999mh}.
He claimed three fundamental conjectures related to the physical meaning of the tachyon and identifying it 
as an instability mode leading to condensation.

The first conjecture claims that the tachyon effective potential has a locally stable minimum whose energy
precisely cancel the D25 brane tension. After condensation, the D25 brane is eaten by the new vacuum.

The second conjecture deals with the fate of the lower dimensional branes. It claims that they can be identified
with solitonic lump-like solutions of string field theory in the background of the D25 brane. The energy of these 
solitonic solutions is conjectured to match the lower dimensional brane tensions~\cite{Recknagel:1998ih,Sen:1999mh}.

These conjectures naturally lead to the third one claiming that the stable vacuum can be identified with the
closed string vacuum with no open string states, in particular D-branes.

These conjectures have been analyzed also in the superstring. Here, we shall not deal with this important issue, and refer 
the interested reader to the recent reviews~\cite{Reviews}.

The check of the first conjecture has been performed by exploiting the level truncation method
first proposed in~\cite{Kostelecky:1988ta}. Several tests have been performed~\cite{Sen:1999nx,Taylor:2000ek,Moeller:2000xv,Beccaria:2003ak}
with very good agreement.

The history of the second conjecture is somewhat less complete. Its foundation lies in previous works 
based on Renormalization Group flows in the first quantized theory~\cite{Fendley:1994rh}. The analysis in the framework of 
string field theory is quite interesting since it attacks the problem in a very explicit way with the hope of 
building the actual lump profile. Also, the method can be extended to superstring theory where the arguments of~\cite{Fendley:1994rh}
should be extended.

Following the initial ideas of~\cite{Harvey:2000tv,deMelloKoch:2000ie}, it has been proposed to study the problem
by expanding  the  lump in a discrete series of modes obtained after  compactification on a circle with radius $R$~\cite{Moeller:2000jy}.
In this way, the second conjecture has been verified at the level of 0.1~\%. A remarkable fact 
is that the compactification radius is a free parameter, a feature not totally exploited in~\cite{Moeller:2000jy}.

The nice results of~\cite{Moeller:2000jy} implicitly show that an extrapolation to infinite level would be rather difficult. 
The calculation is done at level (3,6) with 11 states. This means that all states with level up to 3 are kept in the 
quadratic term of the string field action, while the cubic interaction is computed only for triples of states
with total level up to 6. 
Further work presented the lump potential at level (4,8), but with the focus on the $R\to 1$ limit
where the first tachyon harmonic is marginal and no attempt to improve the check of second conjecture~\cite{Sen:2000hx}.

In this paper, we extend the above calculation pushing it to the level (8, 16) where the typical number of states
is $\sim 300$. Apart from this {\em brute force} calculation, we also explore the dependence on the 
free parameter $R$ and its interplay with the convergence issue of the level expansion. 

We discover that there is an optimal radius $R^* \simeq \sqrt{1.34}$ (in units of $\sqrt{\alpha'}$) where the subleading 
correction $1/L^2$ to the level truncation error is quite small, if not vanishing. At this special point
the convergence of the level expansion is rather smooth and dominated by the 
leading $1/L$ term. Due to the improved scaling behavior, we are able to extrapolate safely our moderately extended calculation and 
check the second conjecture with an accuracy at the level of $10^{-5}$. 

As a byproduct of our exploration in the $R$ variable, we also try to extrapolate the value of the 
marginal tachyon field in the limit $R\to 1$. We present some results on this important issue discussing
in particular the extrapolation to infinite level. These results should be useful in a future extension of the 
analysis and further conjectures of~\cite{Sen:2004cq}.

\section{Basic Definitions and calculation setup}
\label{Sec:Definitions}

We follow~\cite{Moeller:2000jy,Sen:1999xm} for the general framework. Space time is factored as
\be
{\cal M}_{23,0}\times {\cal B}_{2,1},
\ee
where ${\cal M}_{23,0}$ is a purely euclidean spectator space and ${\cal B}_{2,1}$ is the relevant $2+1$ dimensional 
space relevant to the calculation. From the point of view of ${\cal B}_{2,1}$ we study the D1 brane condensation and 
the representation of the D0 brane as a lump. We take space time coordinates $(X, X^0, Y)$ on ${\cal B}$, the two fields
$X$, $X^0$ have Neumann boundary conditions and $X$ is wrapped on a circle of radius $R$. The role of the second spatial 
coordinate $Y$ is just that of providing the brane a direction to move.

The conformal field theory that enter the calculation is 
\be
\mbox{CFT}(X)\oplus \mbox{CFT}',
\ee
where $\mbox{CFT}(X)$ with central charge 1 is relative to the field $X$ and $\mbox{CFT}'$ is a $c=25$ conformal field theory describing the 
other degrees of freedom. This part if fully universal in the calculation in the sense that it appears only through 
Virasoro operators $L_n'$. We complete the theory by adding the ghosts.

About $\mbox{CFT}(X)$, it is possible to maintain a certain amount of universality exploiting the Virasoro $L^X_n$. However, this leads to 
some complications due to possible null states. We prefer to work exclusively with $X$ oscillators $\alpha_n$, a point at variance
with respect to~\cite{Moeller:2000jy} providing a partial cross check of those results at low levels.

In this general framework, the D1 brane tension ${\cal T}_1$ is 
\be
2\pi R\ {\cal T}_1 = \frac{1}{2\pi^2 g_o^2},
\ee
where $g_o$ is the open string coupling. The potential energy of a string field configuration $\Psi$
is
\be
\mbox{potential energy} = \frac{1}{g_0^2} V(\Psi) = 2\pi R \ {\cal T}_1\ 2\pi^2 V(\Psi), 
\ee
where 
\be
V(\Psi) = \frac 1 2 \langle \Psi , Q \Psi\rangle + \frac 1 3 \langle \Psi, \Psi\star\Psi\rangle.
\ee
The potential $V$ is the standard Witten action for the open string field theory~\cite{Witten:1985cc}.
The bracket $\langle \cdot, \cdot\rangle$ denotes the BPZ scalar product. $Q$ is the BRST charge and for the 
string field we adopt the Feynman Siegel gauge fixing. 

The mass of the D1 brane is $M_1 = 2\pi R {\cal T}_1$. Adding the potential energy, we find the total energy
\be
E(\Psi) = 2\pi R{\cal T}_1 (1+2\pi^2 V(\Psi)).
\ee
Since the tension of the D0 brane is ${\cal T}_0 = 2\pi {\cal T}_1$, we find the following form of
Sen's second conjecture
\be
\label{eq:r}
r = \frac{E(\Psi_{\rm lump})}{{\cal T}_0} = R(1+2\pi^2 V(\Psi_{\rm lump})) = 1.
\ee
In principle, we know that for the vacuum solution $\Psi_{\rm vacuum}$, Sen's first conjecture predicts 
$2\pi^2 V(\Psi_{\rm vacuum}) = -1$. Hence, the previous relation can also  be written
\be
r = 2\pi^2 R (V(\Psi_{\rm lump})-V(\Psi_{\rm vacuum})) = 1.
\ee
This second form differs from the former at finite level where also $V(\Psi_{\rm vacuum})$ is computed at a certain level.
It is more accurate at low levels, but rather erratic as the level increases. We shall not exploit it and analyze instead
the better behaved quantity $r$ defined at each level as in Eq.~(\ref{eq:r}).

\subsection{Hilbert space}

The relevant Hilbert space is the $\mathbb{R}$-linear span of the elementary states
\be
s^\alpha s' s^{gh} |n\rangle_\pm,
\ee
where
\be
s^\alpha = \cdots \alpha_{-l^\alpha_3}\ \alpha_{-l^\alpha_2}\ \alpha_{-l^\alpha_1}, \qquad 1 \le l^\alpha_1 \le l^\alpha_2 \le \cdots ,
\ee
\be
s' = \cdots L'_{-l'_3}\ L'_{-l'_2}\ L'_{-l'_1}, \qquad 2 \le l'_1 \le l'_2 \le \cdots ,
\ee
and 
\be
s^{gh} = \cdots b_{-l^b_3}\ b_{-l^b_2}\ b_{-l^b_1}\ \cdots c_{-l^c_3}\ c_{-l^c_2}\ c_{-l^c_1}\ c_1 ,\quad 
1 \le l^b_1 < l^b_2 < \cdots,\  1 \le l^c_1 < l^c_2 < \cdots  .
\ee
The state $ |n\rangle_\pm$ is defined as 
\be
|n\rangle_\pm = \frac 1 2 \left(\exp\frac{inX(0)}{R}\pm\exp\frac{-inX(0)}{R}\right)|0\rangle,
\ee
and notice the relation ($\varepsilon\in\mathbb{Z}_2$)
\be
\alpha_0 |n\rangle_\varepsilon = \frac{\sqrt{2} n}{R}|n\rangle_{-\varepsilon}.
\ee
Of course, the modes with $n>0$ provide the desired $X$ dependence needed in order to construct a non
trivial lump.

We still have to impose a certain set of conditions on the relevant string fields for our 
calculation. Let us define
\be
N^\alpha = \sum_i l^\alpha_i,\quad N' = \sum_i l'_i,\quad N^b = \sum_i l^b_i,\quad N^c = \sum_i l^c_i,
\ee
and the full level $N = N^\alpha + N' + N^b + N^c$, not including the winding contribution. 
The additional conditions are
\begin{enumerate}
\item Twist symmetry: $N$ even, as discussed for instance in~\cite{Gaberdiel:1997mg};
\item $X$ parity: $\varepsilon\cdot (-1)^{N^\alpha} = +1$;
\item ghost number 1: \#b+1 = \#c, where \#b and \#c denote the number of b and c ghosts.
\end{enumerate}

We now give an explicit example of this construction by considering the string field at the generic value $R^2 = 1.3$
which is in the typical range that we shall discuss in the next Sections.
The set of states satisfying the above requirements and with level $L = N+n^2/R^2\le 4$, including now 
the winding contribution from the state $|n\rangle_\pm$, is 
\ba
\psi_{1, n} &=&  c_1|n\rangle_+, \quad  n = 0, 1, 2, \\
\psi_{2, n} &=& b_{-1} c_{-1} c_1|n\rangle_+, \quad n = 0, 1, \\
\psi_{3, n} &=&  {L'}_{-2}  c_1 |n\rangle_+, \quad n = 0, 1, \\
\psi_4 &=& \alpha_{-1} \alpha_{-1}   c_1 |0\rangle_+, \\
\psi_5 &=&  \alpha_{-2}  c_1 |1\rangle_-, \\
\psi_6 &=&  \alpha_{-1} \alpha_{-1}   c_1|1\rangle_+, \\
\psi_7 &=&      b_{-1} c_{-3} c_1 |0\rangle_+, \\
\psi_8 &=&      b_{-2} c_{-2} c_1 |0\rangle_+, \\
\psi_9 &=&      b_{-3} c_{-1} c_1 |0\rangle_+, \\
\psi_{10} &=&  {L'}_{-2} b_{-1} c_{-1} c_1 |0\rangle_+, \\
\psi_{11} &=&  {L'}_{-4}  c_1 |0\rangle_+, \\
\psi_{12} &=&  {L'}_{-2} {L'}_{-2}  c_1 |0\rangle_+, \\
\psi_{13} &=& \alpha_{-1} \alpha_{-1}  b_{-1}, c_{-1} c_1 |0\rangle_+, \\
\psi_{14} &=& \alpha_{-1} \alpha_{-1} {L'}_{-2}  c_1 |0\rangle_+, \\
\psi_{15} &=& \alpha_{-3} \alpha_{-1}   c_1 |0\rangle_+, \\
\psi_{16} &=& \alpha_{-2} \alpha_{-2}   c_1 |0\rangle_+, \\
\psi_{17} &=& \alpha_{-1} \alpha_{-1}, \alpha_{-1}, \alpha_{-1}   c_1 |0\rangle_+ .
\ea
Notice that in our counting of levels, we do not include the operator $c_1$ which is always
present. Also, the state $\psi_5$ is odd under $X\to -X$ and therefore it is built with 
the odd state $|n=1\rangle_-$.

\subsection{Evaluation of the Witten action}

The Witten action can be written in the Feynman-Siegel gauge as
\be
S = \frac{1}{2} \mbox{bpz}(\Psi)\ L_0\ \Psi + \frac{1}{3} \langle V_3 | \Psi^{(1)}\otimes \Psi^{(2)}\otimes \Psi^{(3)}.
\ee
The vertex $\langle V_3 |$ is the cubic vertex coupling the three strings state $\Psi^{(1)}\otimes \Psi^{(2)}\otimes \Psi^{(3)}$. 
We shall provide later precise rules to evaluate its contribution.
Notice that twist symmetry implies that the cubic vertex is symmetric in the three arguments, not just cyclically symmetric. Also 
three elementary states can be coupled if the combination has a projection on the zero momentum sector, i.e. the winding numbers
$(n_1, n_2, n_3)$ satisfy
\be
n_1 + n_2 + n_3 = 0\ \mbox{mod}\ 2 .
\ee

\subsubsection{BPZ conjugation}

The rules to evaluate the BPZ conjugate of a generic state are well known~\cite{Reviews}. Here, for completeness we summarize the recipe. 
Of course, BPZ is linear, so we simply need its action on the elementary states. We have
\be
\mbox{bpz}\{\varphi_1\cdots\varphi_N |n\rangle_\varepsilon \} = (-1)^{\frac 1 2 g(g-1)}\ \varepsilon\langle n|\widetilde{\varphi_N}\cdots\widetilde{\varphi_1},
\ee
where $g = \#b + \#c$, and
\be
\widetilde{\alpha_n} = (-1)^{n+1}\alpha_{-n}, \quad
\widetilde{L'_n} = (-1)^{n}L'_{-n}, \quad
\widetilde{b_n} = (-1)^{n} b_{-n}, \quad
\widetilde{c_n} = (-1)^{n-1} c_{-n}.
\ee

\subsubsection{Kinetic terms}

The kinetic term can only couple states with the same winding $n$ and with the same level $L$. For two such states we 
perform the $\alpha$, $L'$, $b$, $c$ algebra (notice that no $\alpha_0$ can arise) and we are left with the overall
factor depending on the winding and the $X\to -X$ parity of the basic state $|n\rangle_\varepsilon$
\ba
\lefteqn{\frac 1 2 \cdot\frac 1 2 (\langle -n| + \varepsilon_1\langle n|)L_0 (|n\rangle + \varepsilon_2 |-n\rangle) = }&& \\
&=& \frac{1}{4}\left(L-1+\frac{n^2}{R^2}\right)\ \varepsilon_1\ [1+\varepsilon_1\varepsilon_2 + (\varepsilon_1+\varepsilon_2)\delta_{n,0}].
\ea

\subsection{Cubic vertex: conservation rules}

The evaluation of the cubic interaction is conveniently done by means of the conservation rules discussed in~\cite{Rastelli:2000iu}.
That paper deals with all fields we are interested in. These are the b, c ghosts, the $L'$ Virasoro and the primary $\partial X$, i.e. the $\alpha$ oscillators.
Of course, the conservation rules for the $\alpha$ oscillators are those for an anomaly free current, see~\cite{Rastelli:2000iu}, Section 4.2.  

The conservation rules permit to apply a negatively modded field on the cubic vertex. The result is an infinite series of 
$\alpha_0$ and positively modded
operators applied to the three string state $\Psi^{(1)}\otimes \Psi^{(2)}\otimes \Psi^{(3)}$. 
Due to the fact that we work at a fixed order in the level expansion, the infinite series
truncate and we obtain a closed result just by repeatedly applying the operator algebra as well as the conservation laws.

The basic ingredients of the algorithm are certain meromorphic functions with various 
conformal transformation properties (i.e. purely vector, quadratic differentials, or scalar). 
We do not repeat the full discussion in~\cite{Rastelli:2000iu} where all the relevant definitions can be found.
Appendix B of~\cite{Beccaria:2003ak} provide expressions to evaluate the conservation rules quickly in the $\mbox{CFT}'$ and ghost sectors. 
Here we  provide the analogous result for the current $\partial X$.
As discussed in Section 4.2 of~\cite{Rastelli:2000iu}, the conservation laws in the $\alpha$ sector are 
determined by a sequence of meromorphic scalar functions  $f_n(z)$. The general form of $f_n(z)$ is 
\be
f_n(z) = \frac{Z_n(z)}{z^n} ,
\ee
where $Z_n(z)$ is a polynomial. An explicit form of $f_n(z)$ for all $n$ is obtained as follows. Let us define
for each $n$, the sequence $c_{n,n}, \dots, c_{n, 1}$ as 
\be
\frac{1}{z^2-3}\frac{1}{[\tan(3/2\ \arctan z)]^n} = \frac{c_{n,n}}{z^n}+\cdots + \frac{c_{n,1}}{z} + \cdots,
\ee
and
\be
{\cal F}_n = \frac{c_{n,n}}{z^n}+\cdots + \frac{c_{n,1}}{z} .
\ee
Then we have simply
\be
\label{eq:alpha}
f_n(z) = \left\{\begin{array}{ll}
\mbox{pole part of}\ [(z^2-3){\cal F}_n] & : n\ \mbox{odd}  \\
(z^2-3)\ {\cal F}_n & : n\ \mbox{even} 
\end{array}\right.
\ee
Evaluation on Eq.~(\ref{eq:alpha}) reproduces immediately the results Eq.~(4.17) in~\cite{Rastelli:2000iu}.

After the repeated use of the conservation laws, the full cubic interaction between three elementary states 
is reduced to the evaluation of the basic coupling
\be
V_3(n_1, \varepsilon_1; n_2, \varepsilon_2; n_3, \varepsilon_3) = 
\langle V_3 | c_1^{(1)}|n_1\rangle_{\varepsilon_1} \otimes c_1^{(2)}|n_2\rangle_{\varepsilon_2} \otimes c_1^{(3)}|n_3\rangle_{\varepsilon_3} .
\ee
From the conformal field theory definition of the three string vertex, we obtain the following  
explicit value of this quantity 
\ba
\lefteqn{V_3(n_1, \varepsilon_1; n_2, \varepsilon_2; n_3, \varepsilon_3) = 
\frac{1}{8} K^{3-\frac{1}{R^2}(n_1^2+n_2^2+n_3^2)}(1+\varepsilon_1\varepsilon_2\varepsilon_3) \times } && \\
&& (\delta_{n_1+n_2+n_3,0} + 
\varepsilon_1\ \delta_{-n_1+n_2+n_3,0} + 
\varepsilon_2\ \delta_{n_1-n_2+n_3,0} + 
\varepsilon_3\ \delta_{n_1+n_2-n_3,0}),\nonumber
\ea
where $K = 3\sqrt{3}/4$.

\section{Check of Sen's Second Conjecture}
\label{Sec:check2}

We compute the full string field potential at level $(L, 2L)$. This means that we keep all 
fields with level $\le L$ in the kinetic terms and all triples of fields with total level $\le 2L$
in the cubic interaction. The level $L$ includes now the contribution $n^2/R^2$ from the winded vacuum
$|n\rangle_\pm$. In principle there are many possibilities for the level $L$, not necessarily integer due to the $R$ dependence
However, in the following we shall consider only data with integer $L$ since these are the ones leading to a smoother behavior,
as we shall discuss.

We work up to level (8, 16) with various choices for $R^2$, that we use as free parameter. Notice that the full tachyon 
potential indeed is a function of $R^2$ (due to the action of an even number of $\alpha_0$ on the various vacua $|n\rangle_\pm$).

We report in Table~(\ref{tab:1}) the value of $r$ for various levels and radii. 

\TABLE{
\begin{tabular}{||l|cccc|ccc|| }
\hline
$R^2$ & L=2 & 4 & 6 & 8 & 3 & 5 & 7 \\
\hline
1.1 & 1.0250974 & 1.0168271 & 1.0127142 & 1.0102891 & 1.02305   & 1.0157092 & 1.0120622 \\
1.2 & 1.0348878 & 1.020224  & 1.0141754 & 1.0109664 & 1.0305817 & 1.0184029 & 1.0132774 \\
1.3 & 1.0415691 & 1.0218355 & 1.0145796 & 1.0109855 & 1.0355189 & 1.019546  & 1.0135689 \\
1.4 & 1.0478705 & 1.0227330 & 1.0148347 & 1.0110258 & 1.0320652 & 1.018971  & 1.013272  \\
2.0 & 1.0666106 & 1.0285166 & 1.0176278 & 1.0126815 & 1.0465622 & 1.0240124 & 1.0159179 \\
\hline
\end{tabular}

\caption{Ratio $r$ computed at various different radii $R$ and levels $L$. The approximation in the evaluation of the Witten
action is always $(L,2L)$.}

\label{tab:1}
}
In Table~(\ref{tab:1}) we have separated the results with even $L$ from those with odd $L$. The reason can be understood by 
looking at the following Fig.~(\ref{fig:1}) where we have shown the behavior of $r$ at four representative values of $R^2$.

\FIGURE{\epsfig{file=r1.four.eps,width=14cm} 
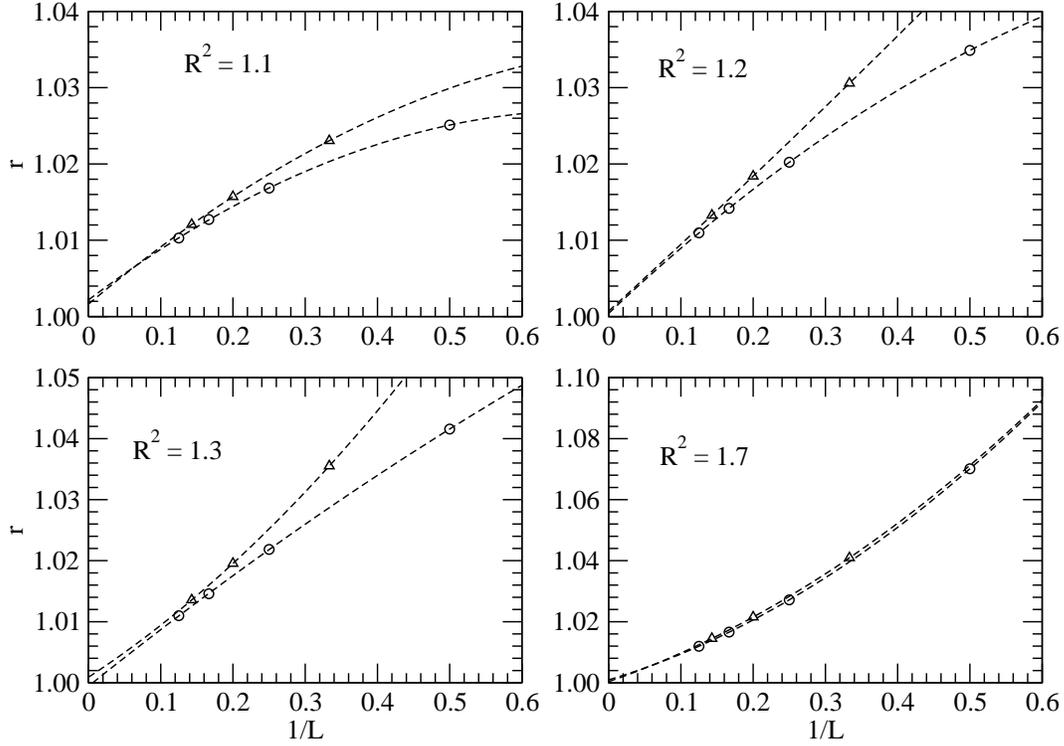
        \caption{Convergence of the ratio as $L\to\infty$ for four representative values of the compactification
                 radius. Circles and triangles denote the even and odd values of $L$ respectively.}
	\label{fig:1}}
It is clear that the two subsequences with even or odd $L$ belong to different smooth curves and 
any extrapolation procedure to the $L\to\infty$ limit must take this fact into account. In particular, we shall attempt
a polynomial extrapolation in the variable $1/L$ on the two separate curves. Since we have a small number of points, we 
simply use a quadratic fit of the form 
\be
r^\pm(L, R) = r^\pm_0(R) + \frac{r^\pm_1(R)}{L} + \frac{r^\pm_2(R)}{L^2},
\ee
where $+(-)$ refers to the subsequence with even (odd) $L$. The results of such a fit are shown in Table~(\ref{tab:2}).
\TABLE{
\begin{tabular}{||l|ccc|ccc|| }
\hline
$R^2$ & $r^+_0$ & $r^+_1$ & $r^+_2$ & $r^-_0$ & $r^-_1$ & $r^-_2$ \\
\hline
1.1  & 1.00218  & 0.0715072 & -0.051347   & 1.00163   & 0.0796027 & -0.0460251  \\
1.15 & 1.00117  & 0.0834426 & -0.0486889  & 1.00099   & 0.086692  & -0.0218726  \\
1.2  & 1.00042  & 0.0894727 & -0.0410577  & 1.00071   & 0.0867362 & 0.00863399  \\
1.25 & 0.999862 & 0.0927268 & -0.0314023  & 1.0007    & 0.083104  & 0.0427123  \\
1.3  & 0.999438 & 0.0946927 & -0.020853   & 1.00091   & 0.0772442 & 0.0797859  \\
1.34 & {\bf 0.999939 } & 0.0880721 & {\bf 0.000373569} & 0.999099  & 0.100666  & -0.0158489  \\
1.35 & 0.999921 & 0.0879025 & 0.00329091  & 0.999045  & 0.100956  & -0.0144504  \\
1.4  & 0.99988  & 0.0867754 & 0.018415    & 0.998795  & 0.102482  & -0.00801743  \\
2    & 0.998946 & 0.101013  & 0.0686399   & 0.999802  & 0.0922072 & 0.144219 \\
\hline
\end{tabular}

\caption{Fit of the ratio $r$ with a quadratic function of $1/L$. The fit is done separately on the even and odd subsequences.}

\label{tab:2}
}
The coefficient $r^+_2$ changes sign at about $R = R^* = \sqrt{1.34}$. The change of convexity of the even subsequence can also 
be seen in Fig.~(\ref{fig:1}). It is clear that at this special radius, the subleading correction $1/L^2$ is very small.
Therefore, we can consider the associated value of $r_0^+(R^*)$ as the best estimate for the ratio $r$ at infinite level.
Indeed, we see from the table, that at $R^*$, the obtained estimate $r_0^+(R^*) = 0.999939$ is quite near the theoretical
prediction $r=1$ with a difference at the level of $10^{-5}$.

Similar considerations can be done working on the odd subsequence. Indeed, when $|r^-_2(R)|$ is small, the 
estimate $r^-_0$ is nearer $r=1$. The precision is smaller than in the case of the even subsequence because
there are only 3 point to be fit. 

In conclusion, the above procedure shows the existence of an optimal compactification radius to 
test the second conjecture. For any degree of accuracy in the level expansion, at $R^*$ the subleading 
corrections in $1/L$ are suppressed and an improvement in the extrapolation to $L\to \infty$ is 
assured.

\subsection{The Marginal Tachyon Mode}
\label{Sec:Marginal}

As a byproduct of our calculation, we can try to estimate the $R\to 1$ value of the tachyon first harmonic in the 
lump solution. This is the coefficient $t_1$ of the state $c_1 |1\rangle_+$
in the string field solution. The problem is that now we cannot vary $R$ and we must resort to 
the extrapolation. 
A sample of data for $t_1$ is reported in Table~(\ref{tab:3})
\TABLE{
\begin{tabular}{||l|cccc|| }
\hline
$R^2$ & L=2 & 4 & 6 & 8 \\
\hline
1.10 & 0.29707354 & 0.33624683 & 0.35166239 & 0.35971677 \\
1.15 & 0.32330778 & 0.35715693 & 0.36893321 & 0.37467583 \\
1.20 & 0.33995093 & 0.37021406 & 0.37973651 & 0.38411374 \\
1.25 & 0.35139367 & 0.37925396 & 0.38730232 & 0.39080940 \\
1.30 & 0.35964714 & 0.38590964 & 0.39295642 & 0.39587931 \\
1.35 & 0.36579239 & 0.39143043 & 0.39744044 & 0.39989409 \\
1.40 & 0.37047136 & 0.39547722 & 0.40095305 & 0.40310957 \\
1.70 & 0.38349706 & 0.40818453 & 0.41199685 & 0.41325659 \\
2.00 & 0.39340980 & 0.41116312 & 0.41404162 & 0.41489224 \\
\hline
\hline
$R^2$ & L=3 & 5 & 7 & \\
\hline
1.10 &  0.31126782 & 0.34191381 & 0.35450203 & \\
1.15 &  0.33696675 & 0.36177155 & 0.37101530 & \\
1.20 &  0.35296316 & 0.37411389 & 0.38134968 & \\
1.25 &  0.36390617 & 0.38268707 & 0.38862042 &\\
1.30 &  0.37183202 & 0.38904484 & 0.39408725 &\\
1.35 &  0.38428460 & 0.39469223 & 0.39856274 &\\
1.40 &  0.38915108 & 0.39853680 & 0.40196426 &\\
1.70 &  0.40194383 & 0.41043369 & 0.41261409 &\\
2.00 &  0.40620485 & 0.41309274 & 0.41461733 &\\
\hline
\end{tabular}

\caption{Marginal tachyon field $|t_1|$ computed at various different radii $R$ and levels $L$. The approximation in the evaluation of the Witten
action is always $(L,2L)$.}

\label{tab:3}
}

Again, we show in Fig.~(\ref{fig:2}) the behavior of $t_1$ as a function of $1/L$
at four representative radii, together with a quadratic fit as before 
\be
t_1^\pm(L, R) = t^\pm_{1,0}(R) + \frac{t^\pm_{1,1}(R)}{L} + \frac{t^\pm_{1,2}(R)}{L^2}.
\ee
We have estimated the extrapolated $t_1(R)$ and its theoretical error as
\be
\label{eq:extr}
\frac{1}{2}(t^+_{1,0} + t^-_{1,0}) \pm \frac{1}{2}(t^+_{1,0} - t^-_{1,0}).
\ee
The result is shown in Fig.~(\ref{fig:3}). We have performed a fit of $t_1(R^2)$ of the form 
$(a_0 + a_1 R^2)/(1+a_2 R^2)$.
Our estimate for $t_1(R=1)$ is $0.351(5)$. 

The quality of the fit is not bad. However, this result depends of course on the choice of the fitting 
function and also on the rather arbitrary choice of the representation Eq.~(\ref{eq:extr}). 
In our opinion, a safe determination of $t_1(R=1)$ would require in our opinion a better accuracy in the level expansion 
as well as additional analytical information about the $t_1$ dependence on $R$ at least in the critical region $R\simeq 1$.

\vskip 1.2cm
\FIGURE{\epsfig{file=t1.four.eps,width=12cm} 
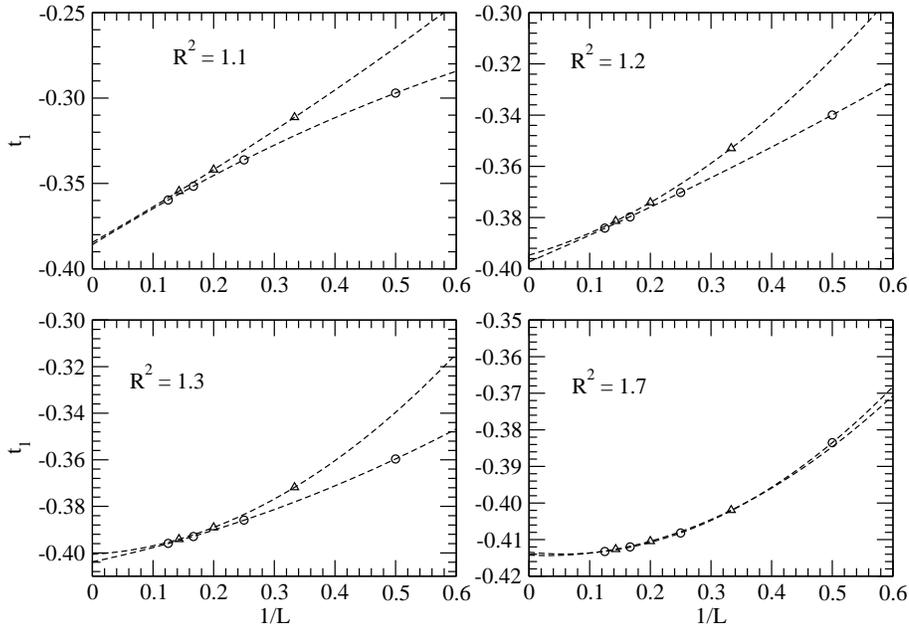
        \caption{Convergence of the marginal tachyon $t_1$ as $L\to\infty$ for four representative values of the compactification
                 radius. Circles and triangles denote the even and odd values of $L$ respectively.}
	\label{fig:2}}

\section{Conclusions}
\label{Sec:Conclusions}

In this brief paper we have tried to improve the available accuracy in the check of Sen's second conjecture
predicting the existence of lump solutions in the background of a Dp-brane and representing a lower 
dimensional D-(p-1) brane. The conjecture asserts that the lump solution energy excess with respect to the 
non perturbative vacuum precisely matches the D-(p-1) brane tension. The ratio of these two quantities is thus
predicted to take the unit value. Previous checks of this prediction confirmed $r=1$ with a precision of about 0.1 \%.

We have extended the level expansion of $r$ up to (8, 16).  In principle the extension could be further 
improved~\footnote{Work is in progress reaching the remarkable (12, 36) level (L. Rastelli, private communication)}.
However, we think that it would be convenient to explore also alternative approaches to the $L\to\infty$
limit. In particular, better extrapolation techniques have been shown to be effective in other similar problems
in the past~\cite{Beccaria:2003ak}.

In this spirit, we have analyzed the compactification radius dependence of the ratio $r$, which is expected to be 
$R$ independent in the infinite $L$ limit. We have shown numerically the existence of an optimal radius $R^* \simeq \sqrt{1.34}$
at which the subleading corrections $1/L^2$ are suppressed. At this special radius we have obtained the estimate 
$r = 0.999939$ from our moderate (8, 16) data. 

Since we explored different radii, it has been natural to analyze the extrapolation of the lump solution in the $R\to 1$
limit. In particular, it is interesting to analyze the first tachyon harmonic which is exactly marginal at $R=1$.
Our best estimate is $t_1(R=1) = 0.351(5)$, but with a rather large theoretical uncertainty.

We hope that this investigation as well as the explicit (8, 16) potential (available on request to the author)
will be useful to clarify the precise relation between the D1~$\to$~D0 marginal transition on a circle and 
the large marginal deformation driven by a Wilson line as investigated in~\cite{Sen:2000hx,Sen:2004cq}.
\vskip 2cm
\FIGURE{\epsfig{file=t1.eps,width=12cm} 
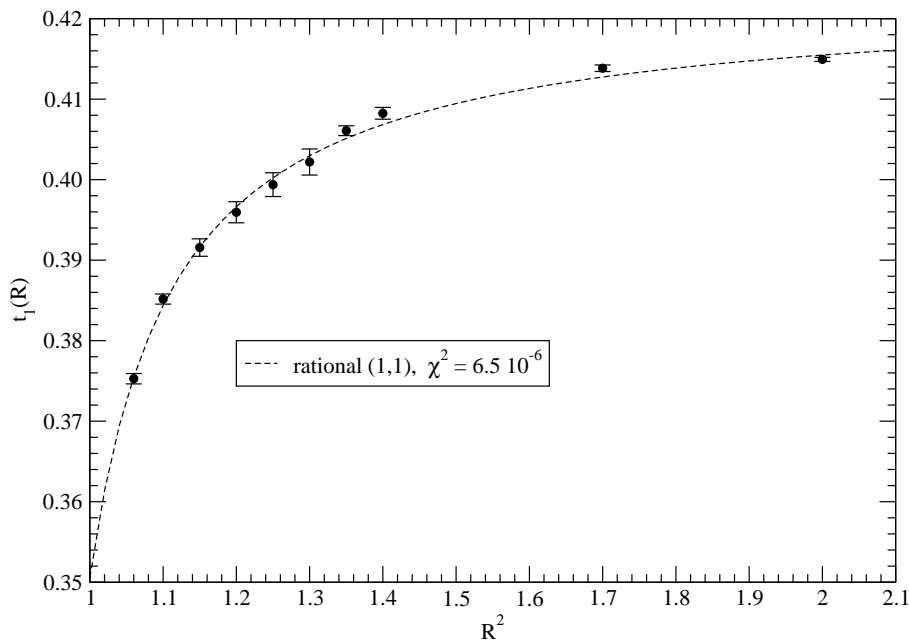
        \caption{Extrapolated marginal tachyon field $t_1(R)$ and rational (1,1) fit.}
	\label{fig:3}}

\end{document}